\documentclass[pre]{revtex4-1}

\usepackage{amsmath}
\usepackage{amssymb}
\usepackage{graphicx}
\usepackage{xcolor}
\newcommand{\nnn}{{\mathcal N}}
\newcommand{\argmin}{
\mathop{\mbox{\rm argmin}
}}
\begin{document}



\title{Rate equation limit for a combinatorial solution of a stochastic aggregation model}


\author{F. Leyvraz}
\email[]{leyvraz@icf.unam.mx}
\altaffiliation{also at Centro Internacional de Ciencias, Cuernavaca, Morelos, M\'exico}
\affiliation{Instituto de Ciencias F\'\i sicas---Universidad Nacional Aut\'onoma de 
 M\'exico, Cuernavaca, Morelos, M\'exico}


\date{\today}

\begin{abstract}
In a recent series of papers, an exact combinatorial solution was claimed for 
a variant of the so-called Marcus--Lushnikov model of aggregation. In this model, a finite
number of aggregates, are initially assumed to be present in the form of monomers. At each time step,
two aggregates are chosen according to certain size-dependent probabilities and irreversibly joined to 
form an aggregate of higher mass. 
The claimed result given an expression for the full probability distribution over all possible size distributions in terms
of the so-called Bell polynomials. In this paper, we develop the asymptotics of this solution in order to check 
whether the exact solution yields correct expressions for the average cluster size distribution as obtained from the Smoluchowski
equations. The answer is surprisingly involved: for the generic case of an arbitrary reaction rate, it is negative, but
for the so-called {\em classical\/} rate kernels, constant, additive and multiplicative, the solutions obtained are indeed exact. 
On the other hand, for the multiplicative kernel, a discrepancy is found in the full solution between the combinatorial solution and the exact
solution. The reasons for this puzzling pattern of agreement and disagreement are unclear. A better understanding of
the combinatorial solution's derivation is needed, the better to understand its range of validity. 

\end{abstract}


\maketitle

\section{Introduction}
\label{sec:intro}
In various systems irreversible aggregation of ``clusters'' 
$A(m)$ of mass $m$  plays an important role. For instance, 
in aerosol physics, suspended particles coagulate (stick) driven by van der 
Waals forces; similar processes are important in polymer chemistry and astrophysics.
The clusters are in general quite varied, going from galaxies to planetary systems. {\color{black}A standard 
reference for aggregation within aerosol physics is the book by Drake \cite{drake}, for systems involving the physics 
of clouds and precipitation, see for instance \cite{rain}.  A broad general introduction
is also given in \cite{KRB}. An overview of related problems that have interested the author
is found in \cite{ley03}.
}

In such systems one is among other things interested in the cluster size 
distribution as a function of time. The simplest approach consists in 
the analysis of kinetic equations. We consider the reaction scheme
\begin{equation}
A(k)+A(l)\mathop{\longrightarrow}_{K(k,l)}\,A(k+l).
\label{eq:1.1}
\end{equation}
The $A(k)$ correspond to aggregates consisting of $k$ monomers, denoted by $A(1)$, and
the numbers $K(k,l)$ denote the {\em rates\/} at which $A(k)$ reacts with $A(l)$. 

The kinetic description of such a system involves the time-dependent
concentrations $c_k(t)$ of $A(k)$. 
The equations read:
\begin{equation}
\dot{c}_j(t)=\frac12\sum_{k,l=1}^\infty K(k,l)c_k(t)c_l(t)\left[
\delta_{k+l,j}-\delta_{k,j}-\delta_{l,j}
\right]
\label{eq:2}
\end{equation}
A basic property of (\ref{eq:2}) is the following: at a formal level, the total mass 
$M_1$ contained in the system
is conserved: 
\begin{equation}
\frac{d}{dt}\sum_{j=1}^\infty jc_j(t)=0
\label{eq:2.1}
\end{equation}
This property can fail, however, if at some finite critical time $t_c$
\begin{equation}
\sum_{k,l=1}^\infty kK(k,l)c_k(t)c_l(t)
\label{eq.2.2}
\end{equation}
diverges. In such a case the total mass starts decreasing with time after this divergence. Such
systems are called gelling systems.

There exist many results concerning these equations. They can be solved exactly 
for the following rate kernels: 
\begin{subequations}
\begin{eqnarray}
K(k,l)&=&1\qquad\mbox{\rm constant kernel}
\label{eq:3a}
\\
K(k,l)&=&k+l\qquad\mbox{\rm additive kernel}
\\
\label{eq:3b}
K(k,l)&=&kl\qquad\mbox{\rm multiplicative kernel}
\label{eq:3c}
\end{eqnarray}
\label{eq:3}
\end{subequations}
The constant and additive kernels are non-gelling, whereas the multiplicative kernel displays gelation. To show
how this may arise, we display its
solution for $c_j(t)=\delta_{j,1}$ 
\begin{subequations}
\begin{eqnarray}
c_j(t)&=&\frac{j^{j-2}}{j!}t^{j-1}e^{-jt}\qquad(t\leq1)
\label{eq:3.1a}
\\
&=&\frac{j^{j-2}e^{-j}}{j!}\frac1t\qquad(t\leq1)
\label{eq:3.1b}
\end{eqnarray}
\label{eq:3.1}
\end{subequations}
For $t\leq1$ it is seen that the mass indeed remains constant. On the other hand, for $t>1$, all 
$c_j(t)$ are decreasing, and the mass does so as well. It is equal to $1/t$. 

Furthermore, an extensive scaling theory exists to describe the large-time and large-size behaviour
of the $c_k(t)$. In the non-gelling cases, there exists a function $s(t)$ known as the characteristic size,
with the following property: as $k\to\infty$ and $t\to\infty$ in such a way that $k/s(t)=x$ remains constant, 
there is a scaling function $\Phi(x)$ such that
\begin{equation}
\lim_{t\to\infty;k/s(t)=x}\left[
k^2c_k(t)\right]
=\Phi(x).
\label{eq:4}
\end{equation}
For details see for instance \cite{ley03}. For gelling systems, a similar behaviour holds in the vicinity
of the critical time:
\begin{equation}
\lim_{t\to t_c;k/s(t)=x}\left[
k^\tau c_k(t)\right]
=\Phi(x).
\label{eq:5}
\end{equation}
where $\tau$ is an exponent that depends on the specific kernel, which is, for instance, equal to $5/2$
for the multiplicative kernel. 

Let us shortly summarise the results of the scaling theory: to this end we define several quantities: the rate constants 
are assumed to be homogeneous in the sizes $k$ and $l$ with exponent $\lambda$, that is
\begin{equation}
K(ak,al)=a^\lambda K(k,l).
\label{eq:5.1}
\end{equation}
We further define the exponents $\mu$ and $\nu$ by the relation
\begin{equation}
K(i,j)\simeq i^\mu j^\nu\qquad(1\ll i\ll j)
\label{eq:5.2}
\end{equation}
{\color{black} describing the behaviour in the {\em strongly asymmetric\/}  case $i\ll j$.}
Both the large-$x$ and small-$x$ behaviours of $\Phi(x)$ are determined by the above exponents: the system is gelling 
if $\lambda>1$, non-gelling otherwise. If the system is non-gelling, the small-$x$ behaviour of $\Phi(x)$ depends on whether 
$\mu>0$, $\mu=0$ or $\mu<0$: in the former case, $\Phi(x)$ goes as $x^{1-\lambda}$, in the third it goes to zero faster than any power,
whereas in the second it has a small-$x$ behaviour that must be determined separately in each particular case. The large-$x$ behaviour on the 
other hand is always of the type $x^{2-\lambda}\exp(-const.\cdot x)$ except when $\nu=1$, in which case the exponent 
arising before the exponential must be determined for every special case. 

This paper concerns a combinatorial solution proposed in \cite{BPC1,BPC2,BPC3,BPC4,BPC5} of a discrete microscopic model 
underlying the Smoluchowski equations (\ref{eq:2}). We shall refer to this solution, which uses in a fundamental manner
the so-called Bell polynomials, as the {\em Bell Polynomial Ansatz} (BPA). 

The structure of the paper is the following: in Section \ref{sec:combsol} we present in detail the discrete Marcus--Lushnikov model
and the BPA; in Section \ref{sec:summary} we summarise the results 
to be shown in the rest of the paper; in Section \ref{sec:asymptotics} we derive the formulae leading to an exact expression for the solution 
of (\ref{eq:2}) as a consequence of the BPA; we also show there that this solution cannot be generally valid; 
the exact validity of the approach for various classical kernels is discussed in Section \ref{sec:classical} and 
we present conclusions in Section \ref{sec:conclusions}.

\section{The discrete Marcus--Lushnikov model and its proposed combinatorial solution}
\label{sec:combsol}

We start by describing a more microscopic model, {\color{black} known
as the Marcus--Lushnikov model}, describing the underlying stochastic dynamics of the aggregation
process \cite{marcus, lushnik1978}. In this model, there are initially $N$ particles, all of which are monomers,
and the states of the system are described by the vector of integers $\underline{n}=(n_1,\ldots,n_N)$ which satisfy the condition
$\sum_{k=1}^\infty k n_k=N$. Here $n_k$ is the total number of aggregates $A(k)$ in the state $\underline{n}$. 

{\color{black}This model, and a purported solution for it, will occupy us throughout this paper. It may therefore be of interest to 
state briefly why this model is of interest. Clearly, it stands with respect to the continuous model of Smoluchowski in the same relation 
as a microscopic model of the molecular dynamics---or more precisely of the Langevin---type might stand to the equations of 
hydrodynamics. And indeed, the same differences exist. In the Monte-Carlo model we study, many questions can be asked which have no meaning
within the hydrodynamical framework: for instance, we may ask by how much the number of monomers at a given time $t$
varies from run to run, or more generally yet, what is its distribution. Such questions have been studied, for instance in \cite{vDongen1, vDongen2}.

Another issue concerns large clusters. The microscopic model under study
only involves clusters of size less than a given number $N$, whereas the Smoluchowski equations involve infinitely many 
possible cluster sizes. When the system gels, it is assumed that part of the mass goes into an ``infinite cluster''. But clearly the 
rate equations can tell us nothing about the nature of such clusters. However, the Marcus--Lushnikov model can be studied for large values
of $N$, and the distribution of cluster sizes can be studies for both ``small'' and ``large'' clusters. Such was the purpose, for instance, of Lushnikov's 
work on the Marcus--Lushnikov model applied to the multiplicative kernel \cite{lushnik1, lushnik2, lushnik3, lushnik4}, where it 
was shown that after gelation there is a single cluster of size of order $N$ accounting for the mass deficit. It should be noted that this is
not necessarily general: it was shown by Monte-Carlo simulations in \cite{brill21} that in certain systems with combined aggregation and
fragmentation, the size of the ``infinite clusters'' grows as $N^\gamma$ with $\gamma<1$. The properties  of such Monte-Carlo simulations 
have also recently been studied in \cite{kali21}. 

From all this follows that there are several properties of interest of aggregating systems---or related generalizations thereof---which cannot be 
obtained from the rate equations alone, and for which the study of stochastic models such as the Marcus--Lushnikov model is of considerable interest. 
}

Let us here make a general remark on notation: in the transition between the above mentioned microscopic model and the 
kinetic model given by (\ref{eq:2}), we must pass to a continuum limit, in which several variables, extensive in the microscopic model, 
are divided by $N$ to yield a continuous variable of (\ref{eq:2}). With the {\em sole exception\/} of the quantities $n_k$, all such 
variables will be denoted by capital letters. 

The aggregation process is then described as a stochastic process in which at each time step a transition takes place between
a state $\underline n$ and another resulting from the aggregation of one aggregate of size $k$ and another of size $l$. Specifically, 
the transition probability between $\underline n$ and $\underline{n}^\prime$ via an aggregation of $A(k)$ and $A(l)$ is only non-zero
if
\begin{subequations}
\begin{eqnarray}
n_j^\prime&=&n_j-1\qquad(j=k,l)
\label{eq:6a}\\
n_{k+l}^\prime&=&n_{k+l}+1
\label{eq:6b}\\
n_j&^\prime=&n_j\qquad(j\neq k,l,k+l)
\label{eq:6c}
\end{eqnarray}
\label{eq:6}
\end{subequations}
The rates  $r_{k,l}$ of this transition is given by
\begin{eqnarray}
r_{k,l}(\underline{n}\to\underline{n}^\prime)&=&K(k,l)n_kn_l\qquad(k\neq l)\nonumber\\
&=&K(k,k)\frac{n_k(n_k-1)}{2}\qquad(k=l)
\label{eq:7}
\end{eqnarray}
Defining the  normalisation  factors $\nnn_k$ and the transition {\em probabilities\/} $p_{k,l}$ as 
\begin{equation}
\nnn_k=\sum_{l}r_{k,l},\qquad p_{k,l}=\nnn_k^{-1}r_{k,l}
\label{eq:8}
\end{equation}
we obtain the dynamics of the probability distribution $P(\underline{n};S)$ by
\begin{equation}
P(\underline{n};S+1)=\sum_{k,l=1}^N\sum_{\underline{n}^\prime}p_{k,l}(\underline{n}^\prime\to\underline{n})P(\underline{n}^\prime;S)
\label{eq:9}
\end{equation}
where $S$ is the time, which in this model only takes integer values. 

An essential difference between this model and the aggregation process described above should be immediately pointed 
out: in this stochastic model, an aggregation event occurs at each time step, whereas in the aggregation process, the $K(k,l)$ 
are {\em rates\/} in a continuous time process. As we shall see, this changes the time variable, but it does not affect anything else. 
The continuous variable equivalent to $S$ is $\sigma=S/N$.

{\color{black} Let us here shortly discuss the somewhat complex issue of the relation between the time $t$ of the original 
Marcus--Lushnikov model as defined by (\ref{eq:10}) and the variable $\sigma$ of the corresponding discrete model.
As noted above, the ``time'' variable is simply $N-K$, where $K$ is the total number of particles at the step $S$. On the other 
hand, an infinitesimal increase of $t$ in (\ref{eq:10}) corresponds to a random variation of the number of particles
depending on the number of reactions taking place in the given time interval. Thus, at fixed $t$, in the continuous time 
Marcus--Lushnikov model, the number of particles is a random variable, whereas for the discrete variant, the number of particles 
is given at each time step. In the limit we are interested in, however, this has no influence, as the variance of the random number of
particles at given time goes to zero as $N\to\infty$. 
}

The original stochastic model introduced by Marcus \cite{marcus} and Lushnikov \cite{lushnik1978} indeed involves
transition rates, and not probabilities: these 
are defined as in (\ref{eq:7}), with no further normalisation; the continuous time equivalent 
of (\ref{eq:9}) is
\begin{eqnarray}
\frac{\partial P(\underline{n};t)}{\partial t}&=&\sum_{k,l=1}^N\sum_{\underline{n}^\prime}\big[
r_{k,l}(\underline{n}^\prime\to\underline{n})P(\underline{n}^\prime;t)- r_{k,l}(\underline{n}\to\underline{n}^\prime)P(\underline{n};t)\big]
\label{eq:10}
\end{eqnarray}
This process is the true equivalent of the reaction rate equations (\ref{eq:2}), but we shall concentrate on the dynamics 
defined by (\ref{eq:9}), since it is for this that a combinatorial solution,
the Bell Polynomial Ansatz (BPA), has been proposed \cite{BPC1,BPC2,BPC3,BPC4,BPC5}.
To emphasize the difference, we shall call the model described by (\ref{eq:9}) the {\em discrete Marcus--Lushnikov model}

To state this solution we define some additional quantities: the sequence $\xi_k$ is defined by the following recursion relation
\begin{subequations}
\begin{eqnarray}
(k-1)\xi_k&=&\sum_{l=1}^{k-1}K(l,k-l)\xi_l\xi_{k-l}
\label{eq:11a}\\
\xi_1&=&1
\label{eq:11b}
\end{eqnarray}
\label{eq:11}
\end{subequations}
Here (\ref{eq:11a}) only applies for $k\geq2$. 
These quantities play an important role in the short-time behaviour of the solutions of (\ref{eq:2}). As $t\to0$
one has for instance, for the initial condition $c_k(0)=\delta_{k,1}$
\begin{equation}
c_k(t)\simeq\xi_kt^{k-1}.
\label{eq:12}
\end{equation}
Their asymptotic behaviour as $k\to\infty$ has been studied \cite{ley03,theta,vDongenLarge} and they behave as $k^{-\lambda}$
except for the case of kernels having $\nu=1$, for which the behaviour must be studied on a case-by-case basis. 

We now introduce some further notation: underlined letters such as $\underline z$ and $\underline w$ will always
represent vectors with components as follows:  $\underline{z}=(z_1,\ldots,z_N)$. We 
shall use the following abbreviated notations involving the vector of complex variables 
\begin{equation}
\underline{z}^{\underline{w}}=\prod_{k=1}^Nz_k^{w_k}\qquad \underline{z}\,\underline{w}=(z_1w_1,\ldots,z_Nw_N)
\label{eq:14}
\end{equation}
We further introduce the following abbreviations
\begin{equation}
\underline{\xi}=(\xi_1,\ldots,\xi_N)\qquad\underline{m}!=(1!,2!,\ldots,N!).
\label{eq:14a}
\end{equation}
We now define the generating function of the probability distribution $P(\underline{n};S)$ as a function
of
\begin{equation}
\Psi(\underline{z};S)=\sum_{\underline{n}}P(\underline{n};S)\underline{z}^{\underline{n}},
\label{eq:13}
\end{equation}
The BPA yields the following expression for $\Psi(\underline{z};S)$
\begin{equation}
\Psi(\underline{z},S)=\frac{B_{N,N-S}(\underline{m}!\underline{\xi}\underline{z})}{B_{N,N-S}(\underline{m}!\underline{\xi})}
\label{eq:15}
\end{equation}
Here $B_{N,N-S}(\underline{x})$ is the {\em Bell polynomial\/} defined as follows: the series of polynomials
$B_{n,m}(\underline{a})$, where $\underline{a}$ is (exceptionally) an {\em infinite\/} sequence $(a_1,a_2,\ldots)$,
are given by the generating function
\begin{equation}
\sum_{n\geq m\geq1}^\infty B_{n,m}(\underline{a})\frac{x^n}{n!}y^m=\exp\left[
y\left(
\sum_{r=1}^\infty \frac{a_r}{r!}x^r
\right)
\right]
\label{eq:16}
\end{equation}
It is, however, readily checked that $B_{n,m}(\underline{a})$ only depends on $a_k$ for $1\leq k\leq n-m+1$, so that the expressions 
given above are well-defined. For more details on Bell polynomials, see for instance \cite{comtet}.

\section{Summary of results}
\label{sec:summary}

The quantities $c_j(t)$ satisfying (\ref{eq:2}) correspond, in terms of the stochastic model described above, to the quantities
\begin{equation}
\overline{n}_j(S)=\sum_{\underline n}n_jP(\underline{n};S).
\label{eq:17}
\end{equation}
Additionally to its $S$ dependence, $\overline{n}_j(S)$ depends on $N$. It is well-known \cite{vDongen1,vDongen2,norris} that, in the
continuum limit, the discrete Marcus--Lushnikov model tends to the kinetic equations in the following sense:
\begin{subequations}
\begin{eqnarray}
\lim_{\substack{S,N\to\infty\\S/N=\sigma}}\frac{\overline{n}_j(S)}{N}&=&c_j(t)
\label{eq:18a}\\
\sum_{k=1}^\infty c_k(t)&=&1-\sigma
\label{eq:18b}
\end{eqnarray}
\label{eq:18}
\end{subequations}
This means that (\ref{eq:18b}) provides the correct connection between $t$ and $\sigma$, and that taking the $N\to\infty$
limit while maintaining $\sigma$ constant leads to the correct concentration profile $c_j(t)$. Additionally, it can be shown that 
not only does the average value of $n_j(s)$ converge to the correct limit, but that the probability distribution becomes infinitely sharp
around the values defined by the kinetic equations (\ref{eq:2}).

Using an asymptotic formula for the Bell polynomials {\color{black} to be derived later, see (\ref{eq:26})}, 
one finds an exact expression for the above described limit in
terms of the BPA (\ref{eq:15}): define
\begin{subequations}
\begin{eqnarray}
G(w)&=&\sum_{j=1}^\infty \xi_jw^j
\label{eq:19a}\\
w^\star(\sigma)&=&\argmin_{0<w<w_c}\big[
(1-\sigma)\ln G(w)-\ln w
\big]
\label{eq:19b}
\end{eqnarray}
\label{eq:19}
\end{subequations}
where $w_c$ is the convergence radius of $G(w)$. One then finds
{\color{black}
\begin{equation}
\lim_{\substack{N,S\to\infty\\S/N=\sigma}}\frac{\overline{n}_j(S)}{N}=\left.z_j\frac{\partial}{\partial z_j}\ln\Psi(\underline{z},S)
\right|_{\underline{z}=\underline{1}}=
\left.z_j\frac{\partial}{\partial z_j}\ln B_{N,N-S}(\underline{m}!\underline{\xi}\underline{z})
\right|_{\underline{z}=\underline{1}}=
\frac{1-\sigma}{G[w^*(\sigma)]}\,\xi_jw^*(\sigma)^j
\label{eq:20}
\end{equation}
Here we have used the definition of $\Psi(\underline{z};S)$, see (\ref{eq:13}), in the first equality and the BPA, see (\ref{eq:15}),
in the second. The third will follow, as mentioned above, from asymptotic estimates on Bell polynomials to be derived 
in Section \ref{sec:asymptotics}, see (\ref{eq:26}).}
From this we deduce that the r.h.s.~of (\ref{eq:20}) is an exact solution of (\ref{eq:2}) whenever (\ref{eq:15})
is a solution of the discrete Marcus--Lushnikov model defined by (\ref{eq:9}), and viceversa, we obtain that the
BPA, see (\ref{eq:15}), cannot be an exact solution of the discrete Marcus--Lushnikov model unless
(\ref{eq:20}) solves the Smoluchowski equations (\ref{eq:2}). 

Generically, it is readily seen from what we know of scaling theory that (\ref{eq:20}) does not solve the Smoluchowski equations.
Indeed, since the entire aggregation process described in the discrete Marcus--Lushnikov model stops when $S=N$, it
follows that the infinite time limit corresponds to the $\sigma\to1$ limit. As is also readily seen, if $G^\prime(w)\to\infty$
as $w\to w_c$, $w^*(\sigma)\to w_c$ as $\sigma\to1$. (\ref{eq:20}) thus leads to the scaling form
\begin{equation}
c_j(t)\simeq j^{-\lambda}\exp\left[
-j\ln\left(
\frac{w^\star(\sigma)}{w_c}
\right)
\right]\left(
\sum_{k=1}^\infty c_k(t)
\right),
\label{eq:21}
\end{equation}
{\color{black} the derivation of which is given in Appendix  \ref{app:a}}.
To describe this in terms of the scaling theory, we set
\begin{equation}
s(t)=\left[
\ln\left(
\frac{w^\star(\sigma)}{w_c}
\right)
\right]^{-1}.
\label{eq:21.01}
\end{equation}
Using as always the normalisation
\begin{equation}
\sum_{k=1}^\infty kc_k(t)=1
\label{eq:21.02}
\end{equation}
we obtain after some straightforward manipulations that, for the non-gelling case, the scaling function
is of the form
\begin{equation}
\Phi(x)=const.\cdot x^{2-\lambda}e^{-x}.
\label{eq:21.03}
\end{equation}
The elementary computations are given in Appendix \ref{app:a}. A particularly problematic feature of this result
is that it holds good for $\lambda<2$. This means in particular that it does not yield gelling behaviour for 
$\lambda>1$.

While this is in good agreement with the large-$j$ behaviour as known from scaling theory, it is in strong disagreement 
with the known small-$j$ behaviour. Note that the discrepancy involves the size distribution, and not the correspondence 
between $t$ and $\sigma$. Further, since the formula does agree
in the limit of sizes large with respect to typical size, the disagreement can also not be resolved by changing the
definition of the sequence $\xi_j$: such an attempt could introduce agreement at small values of $j$ but would then
spoil the agreement at large $j$. 

Specifically, the kernel solved in \cite{BPC5}, $K(k,l)=1/k+1/l$ has, as can be shown exactly, see \cite{ley21}, the
following large time behaviour
\begin{equation}
\lim_{t\to\infty}\left(
\sum_{k=1}^\infty
c_k(t)
\right)^{-1}\exp\left(
\int_0^t\sum_{k=1}^\infty c_k(t^\prime)\,dt^\prime
\right)c_1(t)
=1
\label{21.1}
\end{equation}
Clearly this result is incompatible with (\ref{eq:21}). 

Yet another kind of behaviour that cannot
be subsumed under the scheme of (\ref{eq:21}) is the following: define $\alpha_j$ and $w_j$ via
\begin{equation}
\lim_{t\to\infty}\left[
t^{w_j}c_j(t)
\right]=\alpha_j
\label{eq:21.2}
\end{equation}
Now it can be shown that the $w_j$ in certain cases depend non-trivially on $j$. For instance, for the kernel $K(k,l)=2-q^k-q^l$
with $0<q<1$, which has been extensively treated in \cite{qmodel1, qmodel2}, one finds that
\begin{equation}
w_j=2-q^j.
\label{eq:21.3}
\end{equation}
Similar behaviour also arises in the case in which the reaction rates $K(k,l)$ take 3 different constant values,
depending on whether both masses are equal to one, only one, or neither \cite{mobilia}. 

\begin{figure}
\includegraphics[scale=0.3]{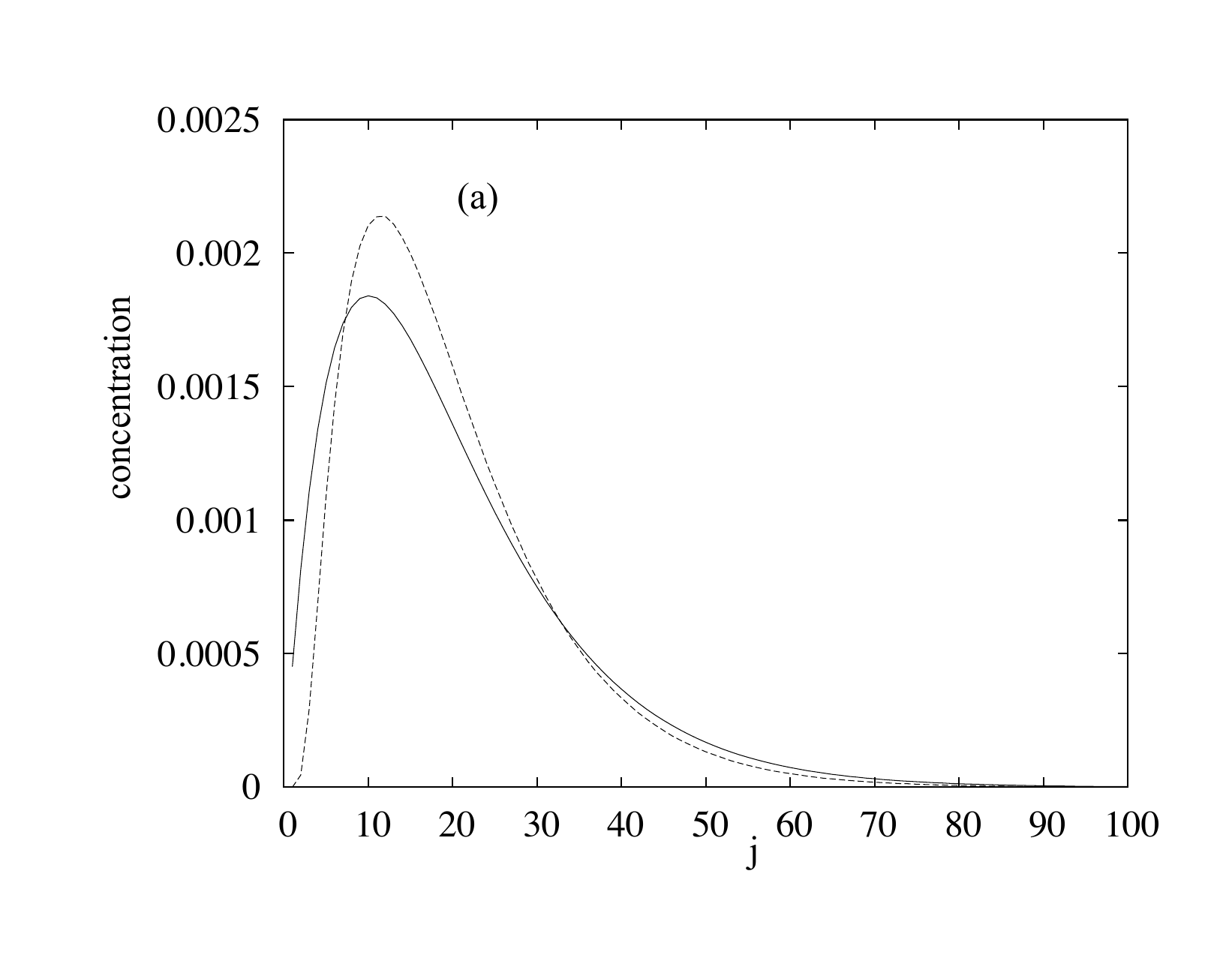}
\includegraphics[scale=0.3]{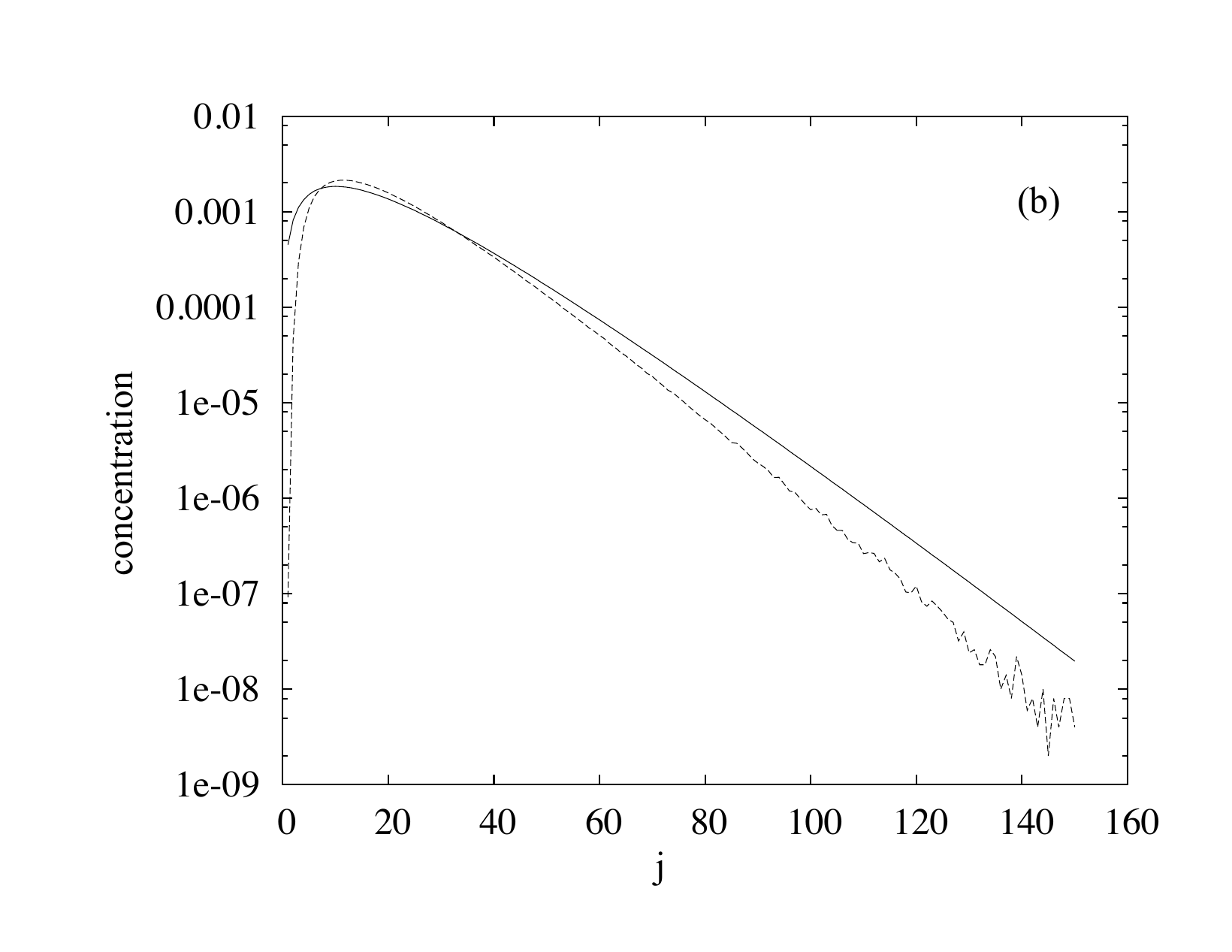}
\includegraphics[scale=0.3]{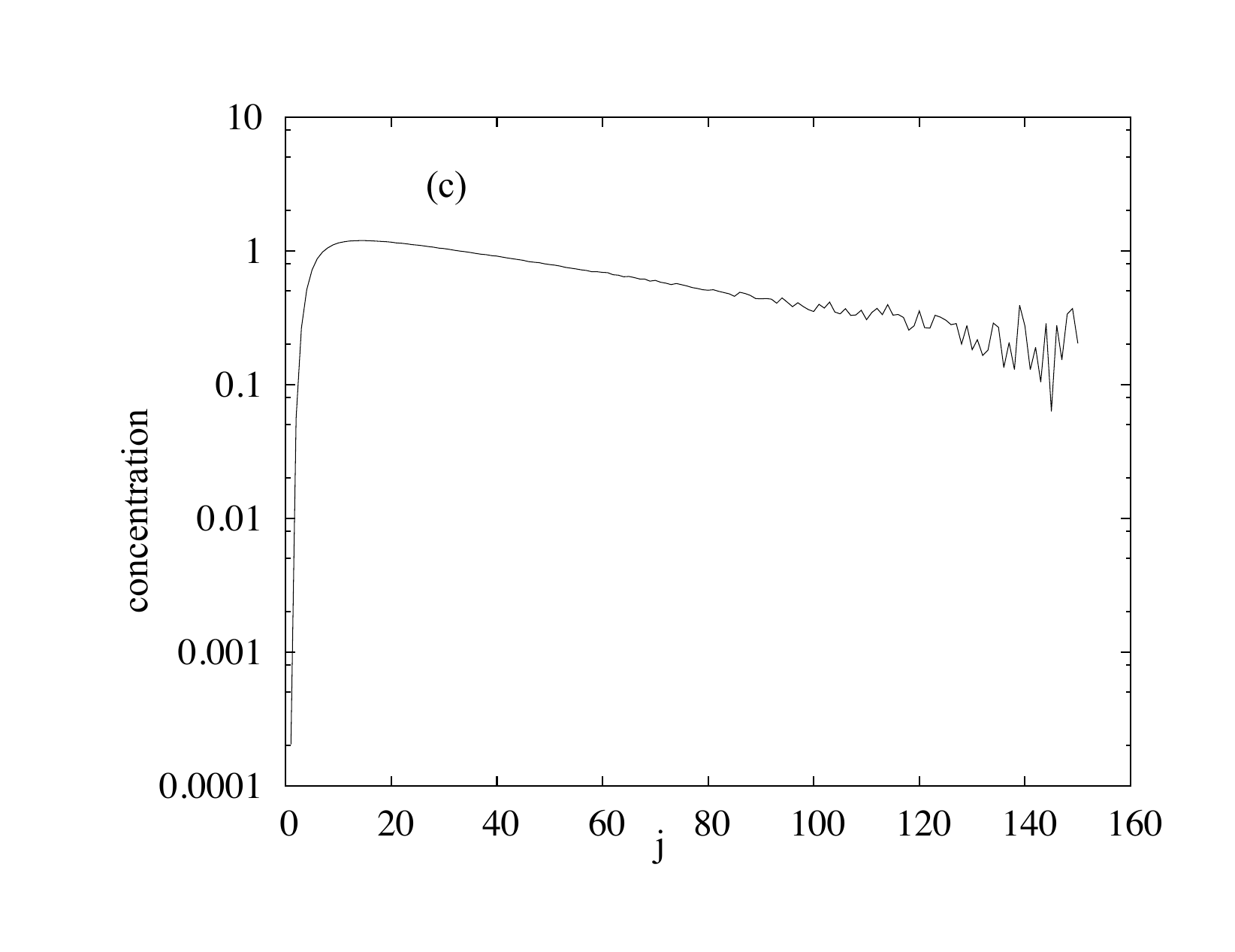}
\includegraphics[scale=0.3]{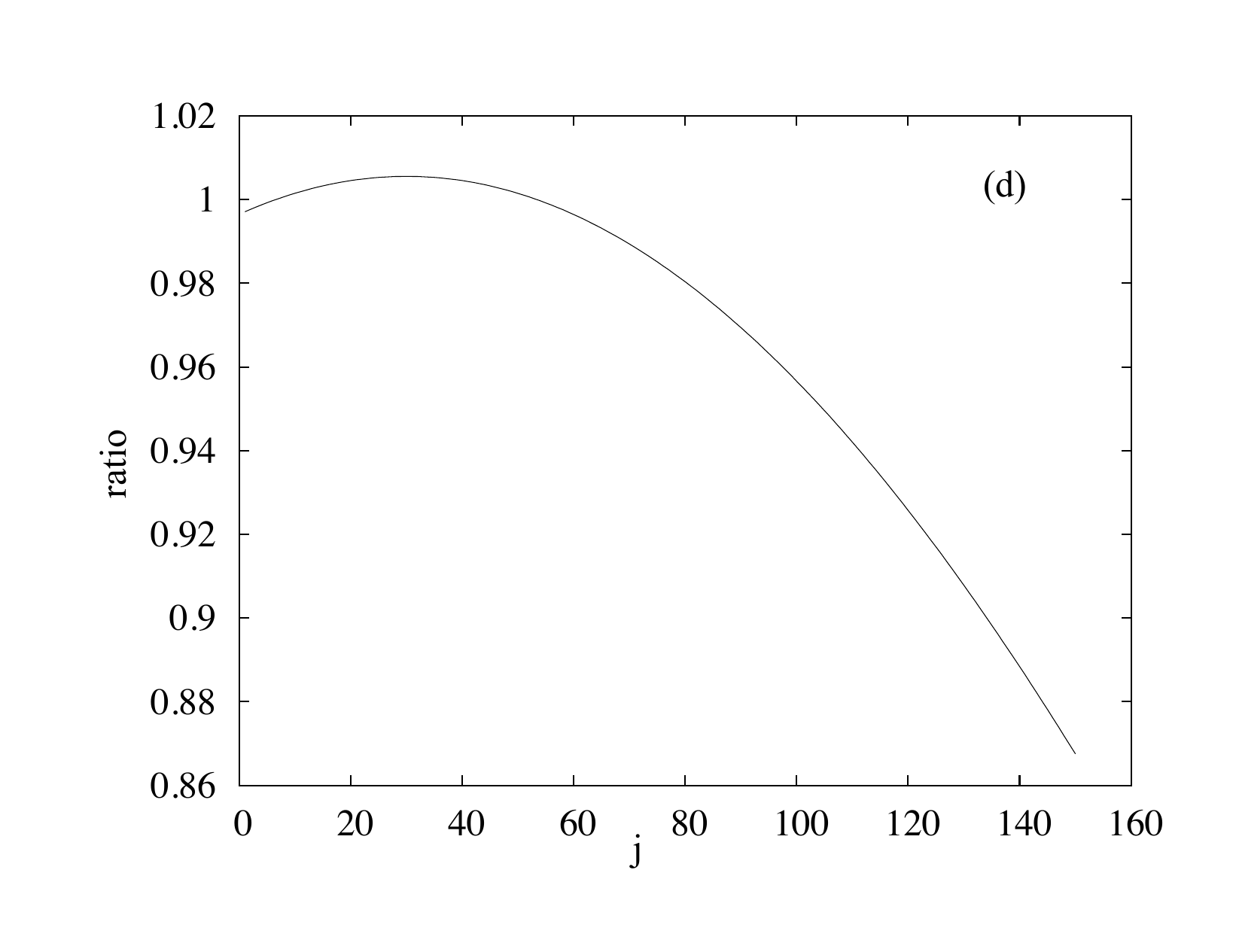}
\caption{
\color{black}Here we compare the cluster size distribution obtained through the BPA to that obtained through direct simulation
of the discrete Marcos--Lushnikov model. The case considered is a system with the reaction rates $K(k,l)=1/k+1/l$,
with $N=5\,000$ and $S=4\,750$, implying a total number of particles of 250. This corresponds to a number density $\sum_kc_k(t)$
of $1/20$, and hence, from the scaling theory of Type III kernels \cite{ley03}, the typical size is of the order of 20.
Since the total number of monomers is $5\,000$, we are not strongly affected by finite size effects. On all figures,
the $x$ axis corresponds to the cluster size. The simulations performed involved $10^5$ runs.
Figure (1a) displays the two cluster size distributions on a linear scale, the full line denoting the BPA and
the dotted line the simulation results. Figure (1b) shows the same on a log-log scale, which 
indicates the very strong deviations at the origin, together with reasonable agreement at large sizes. Figure (1c) tests this 
more precisely by showing the logarithm of the quotient of the numerical solution to the BPA one. There we see how, after very
strong initial variations, the quotient slowly decreases instead of tending to a constant,as predicted here. The reason for the discrepancy appears to 
lie in the non-asymptotic behavior of the BPA expression for the concentration profile, shown in Figure (1d) which displays the quotient of the exact BPA results and its asymptotic behaviour.  
}
\label{fig:1}
\end{figure}

We therefore see that the claim of general validity made for the BPA in \cite{BPC1,BPC2,BPC3,BPC4,BPC5},
cannot be upheld. The reasoning presented in these papers must thus contain some flaw, which is however
by no means obvious. {\color{black} In order to illustrate the nature and the extent of the possible disagreement
between the BPA and the results of a Monte-Carlo simulation of the discrete Marcus--Lushnikov model,
let us consider the specific case illustrated in the Figure \ref{fig:1}, the details of which are given in the caption: as
we see, there is a very strong disagreement for low sizes, specifically for monomers, whereas the two cluster size
distributions are roughly proportional for sufficiently large sizes.}

Nevertheless, the identity between the small-$j$ and the large-$j$ behaviours is
characteristic for the so-called classical kernels, namely the constant, additive, and
multiplicative kernels, as well as their combinations. We shall in fact explicitly show that (\ref{eq:20}) does provide
an exact solution in those cases, for the multiplicative kernel at least before the gel time. 

The fact that the average value of $n_j/N$ is exact for these kernels clearly does not guarantee the accuracy of the full 
probability distribution. An explicit  comparison between the exact distribution of the discrete Marcus--Lushnikov model and the BPA
was performed for the case $N=20$ and $S=4$, starting from the initial condition $n_1=20$ and $n_j=0$ for $j\geq2$. 
The former is readily evaluated  for small systems using explicit enumeration of (\ref{eq:9}). 

These were found to coincide exactly for both the constant and the additive kernel, yet to differ in the case of the multiplicative kernel. 
We state the results for the latter case. At $S=4$, five states only have non-zero probabilities:
\begin{eqnarray}
\underline{n}_1&=&(12,4,\ldots)
\qquad\underline{n}_2=(13,2,1,\ldots)\nonumber\\
\underline{n}_3&=&(14,0,2,\ldots)\qquad
\underline{n}_4=(14,1, 0,1,\ldots)\\
\underline{n}_5&=&(15,0,0,0,1,\ldots)\nonumber
\label{eq:22}
\end{eqnarray}
where the final zeroes have not been written explicitly. The exact probabilities for the multiplicative kernel are given by
\begin{equation}
\begin{array}{ll}
P(\underline{n}_1;4)=\displaystyle{\frac{130}{517}}=0.2515\ldots&P(\underline{n}_2;4)=\displaystyle{\frac{22\,400}{48\,081}}=0.4659\ldots\\
\\
P(\underline{n}_3;4)=\displaystyle{\frac{22\,480}{336\,567}}=0.0668\ldots&P(\underline{n}_4;4)=\displaystyle{\frac{460\,485}{2\,580\,347}}=0.1785\ldots\\
\\
P(\underline{n}_5;4)=\displaystyle{\frac{96\,552}{2\,580\,347}}=0.0374\ldots&{}
\end{array}
\label{eq:22.1}
\end{equation}
whereas the ones arising from the BPA are
\begin{equation}
\begin{array}{ll}
P(\underline{n}_1;4)=\displaystyle{\frac{273}{1\,081}}=0.2525\ldots&P(\underline{n}_2;4)=\displaystyle{\frac{504}{1\,081}}=0.4662\ldots\\
\\
P(\underline{n}_3;4)=\displaystyle{\frac{72}{1\,081}}=0.0666\ldots&P(\underline{n}_4;4)=\displaystyle{\frac{192}{1\,081}}=0.1776\ldots\\
\\
P(\underline{n}_5;4)=\displaystyle{\frac{40}{1\,081}}=0.0370\ldots&{}
\end{array}
\label{eq:22.2}
\end{equation}
%
%
While it is seen that these are numerically quite close, the two are clearly different. On the other hand, for 
the additive and constant kernels, even for the case $N=20$ and $S=16$, which has 70 accessible states, the comparison 
between the BPA and an exact enumeration ,yields perfect agreement. There is thus essentially no doubt concerning the 
exactness of these two solutions. Indeed, Lushnikov \cite{lushnik1978,lushnik2011} has presented exact solutions for these kernels, for the 
corresponding continuous model. These can presumably be carried over to the discrete case. 

\section{The $N\to\infty$ average cluster size distribution}
\label{sec:asymptotics}
Let us first derive an explicit expression for $\overline{n}_j$ as a function of $\Psi(\underline{z};S)$. One finds
\begin{equation}
\overline{n}_j=z_j\left.\frac{\partial}{\partial z_j}\ln\Psi(\underline{z};S)\right|_{\underline{z}=1}
\label{eq:23}
\end{equation}
Substituting the BPA (\ref{eq:15}) into (\ref{eq:23}) yields
\begin{equation}
\overline{n}_j=z_j\left.\frac{\partial}{\partial z_j}\ln\left[
B_{N,N-S}(\underline{m}!\underline{\xi}\underline{z})
\right]
\right|_{\underline{z}=1}
\label{eq:24}
\end{equation}
To evaluate this in the limit of large $N$ and $S$ with $S/N=\sigma$, and $0<\sigma<1$, we need an appropriate 
asymptotic expression for the Bell polynomial. 

{\color{black}From the definition of Bell polynomials, see (\ref{eq:16}), we obtain using Cauchy's expression for the coefficients of
a power series:
\begin{equation}
\sum_{m=1}^\infty \frac1{n!}B_{n,m}(\underline a)y^m=\frac{1}{2\pi i}\oint_C \frac{dw}{w^{n+1}}\,\exp\left[
y\left(
\sum_{r=1}^\infty \frac{a_r}{r!}w^r
\right)
\right]
\label{eq:24.1}
\end{equation}
Here $C$ is a contour enclosing the origin, but no singularity of the integrand.
Now, comparing the coefficients of $y$ one obtains {\color{black}the following well-known relation:}
\begin{equation}
\frac{m!}{n!}B_{n,m}(\underline a)=\frac{1}{2\pi i}\oint_C \frac{dw}{w^{n+1}}\,
\left(
\sum_{r=1}^\infty \frac{a_r}{r!}w^r
\right)^m
\label{eq:24.2}
\end{equation}
A substitution now yields straightforwardly}
\begin{widetext}
\begin{subequations}
\begin{eqnarray}
\frac{(N-S)!}{N!}B_{N,N-S}(\underline{m}!\underline{\xi}\underline{z})&=&
\frac{1}{2\pi i}\oint_C \frac{G(w|\underline{z})^{N-S}}{w^{N+1}}dw
\label{eq:25a}
\\
G(w|\underline{z})&=&\sum_{j=1}^\infty \xi_jz_jw^j
\label{eq:25b}
\end{eqnarray}
\label{eq:25}
\end{subequations}
\end{widetext}
Here $C$ is a contour enclosing the origin, but no singularity of $G(w|\underline{z})$. Specifically, we may think of a circle around the origin 
with a radius $r<w_c$. Since the Taylor series of $G(w)$ has positive coefficients, the maximum value of $G(w)$ on this circle
is assumed as the real axis is crossed. The same remark holds for the integrand of (\ref{eq:25}). 

It therefore follows, using {\color{black}the method of stationary phase}, that
\begin{widetext}
\begin{equation}
\lim_{\substack{S,N\to\infty\\S/N=\sigma}}\frac1N\ln\left[
\frac{(N-S)!}{N!}B_{N,N-S}(\underline{m}!\underline{\xi}\underline{z})
\right]=\left(
1-\sigma
\right)\ln G[w^\star(\sigma|\underline{z})|\underline{z}]-\ln w^\star(\sigma|\underline{z})
\label{eq:26}
\end{equation}
\end{widetext}
where $w^\star(\sigma|\underline{z})$ is defined similarly to (\ref{eq:19b})
\begin{equation}
w^\star(\sigma|\underline{z})=\argmin_{0<w<w_c(\underline{z})}\big[
(1-\sigma)\ln G(w|\underline{z})-\ln w
\big].
\label{eq:27}
\end{equation}
{\color{black}The somewhat messy details are relegated to Appendix \ref{app:b}. Summarizing, however, it may be
said that this is a straightforward application of the method of stationary phase, as described in Bender and Orszag \cite{BO}.}
Note that the asymptotic expression (\ref{eq:26}) does not hold universally for all sequences $\xi_j$. Essential use 
is made both of the positivity of the $\xi_j$ and of the fact that the $\xi_j$ grow regularly (in the sense, for instance, that the nearest
singularity to the origin is on the positive real axis, to the exclusion of other singularities at the same distance). The above
derivation of (\ref{eq:26}) is quite elementary, and I assume it to be already known, but have not found a reference for it. 

We now need to evaluate the derivative with respect to $z_j$ of the r.h.s.~of (\ref{eq:26}). {\color{black} We first use the following well-known 
fact: for an arbitrary function $\phi(x,\lambda)$, let
\begin{equation}
x^\star(\lambda)=\argmin_x(f(x,\lambda))\qquad g(\lambda)=f(x^*(\lambda),\lambda),
\end{equation}
 then 
 \begin{equation}
g^\prime(\lambda)=\frac{\partial f}{\partial\lambda}(x^*(\lambda),\lambda)
\end{equation}
To use this result, we take the function
\begin{equation}
(1-\sigma)\ln G(w(\sigma|\underline{z})|\underline{z})-\ln w(\sigma|\underline{z}).
\end{equation}
as $f(x,\lambda)$, where $w$ plays the role of $x$ and $z_j$ the role of $\lambda$.

Using this result}, we only need to take the derivative with respect to the $\underline{z}$-dependence of $G(w(\sigma)|\underline{z})$. 
This leads to
\begin{equation}
\lim_{\substack{S,N\to\infty\\S/N=\sigma}}\frac{\overline{n}_j}{N}=\frac{1-\sigma}{G[w(\sigma)]}\,\xi_jw(\sigma)^j
\label{eq:28}
\end{equation}
as stated earlier. 

Let us now fill in some of the missing details: we first show that, on the interval $0\leq w\leq w_c$ the function
$(1-\sigma)\ln G(w)-\ln w$ indeed takes a minimum at $w(\sigma)$. Indeed, the second derivative is given by
\begin{equation}
\frac{1-\sigma}{G(w)^2}
\left[G(w)G^{\prime\prime}(w)
-G^{\prime}(w)^2
\right]+\frac1{w^2}>0.
\label{eq:29}
\end{equation}
where the positivity follows from the positivity of the $\xi_j$ and the Cauchy--Schwarz inequality. 
As a consequence, because of the Cauchy--Riemann equations, the same function has a local maximum
in the imaginary direction, which is the one of the contour. 

We now discuss under what circumstances $w(\sigma)$ is an interior minimum
satisfying the equation
\begin{equation}
(1-\sigma)w(\sigma)\frac{G^{\prime}[w(\sigma)]}{G[w(\sigma)]}=1.
\label{eq:30}
\end{equation}
As we move from $w=0$ to $w=w_c$, both $G^\prime(w)$ and $G(w)$
increase. If as $w\to w_c$ $G^\prime(w)$ diverges, then the existence of a (unique)
solution of (\ref{eq:30}) follows straightforwardly. On the other hand, if $G^\prime(w_c)<\infty$,
then there is a value $0<\sigma<1$ such that (\ref{eq:30}) has no interior solution. As we shall see, this
is what happens in the case of the post-gel solution of the multiplicative kernel.

\section{Exact validity for the classical kernels}
\label{sec:classical}

Let us now check that for the three classical cases, the solution given by (\ref{eq:20}) indeed corresponds to 
the known exact solution. In all cases, the quantities $\xi_j$ defined in (\ref{eq:11}) and the corresponding 
generating function $G(w)$ defined in (\ref{eq:19a}) are known exactly, so that verifying the correctness of
the solution (\ref{eq:20}) is in principle straightforward.

Let us consider the case of the multiplicative kernel, since the average $n_j$ for
the other two can be expressed in terms of binomial coefficients, for which the results are readily obtained.

{\color{black} In the following we perform the computations reqired to evaluate the
BPA solution, but all steps are part of the standard 
approach to solving the multiplicative kernel, as reviewed, for instance, in \cite{ley03}}.
It follows from the recursion relation (\ref{eq:11}), that the generating function for the $\xi_j$, namely
$G(w)$, satisfies the ordinary differential equation: 
\begin{equation}
wG^\prime(w)-G(w)=\frac12\left[
wG^\prime(w)
\right]^2\qquad G^\prime(0)=1
\label{eq:31}
\end{equation}
This has the unique implicit solution
\begin{equation}
w=\left(1-\sqrt{1-2G}\right)
\exp\left(\sqrt{1-2 G }-1\right)
\label{eq:32}
\end{equation}
We now substitute (\ref{eq:32}) into
\begin{equation}
(1-\sigma)\ln G-\ln w
\label{eq:33}
\end{equation}
and minimize with respect to $G$. One obtains
\begin{equation}
G=2\sigma(1-\sigma)\qquad w=2\sigma e^{-2\sigma}
\label{eq:34}
\end{equation}
If we finally substitute these expressions into the r.h.s.~of (\ref{eq:20}), one obtains
\begin{equation}
c_j(t)=\xi_j(2\sigma)^{j-1}e^{-2j\sigma}.
\label{eq:35}
\end{equation}
Finally, we need to connect $\sigma$ and $t$. We know the expression (\ref{eq:18b}) for $\sigma$. But
we can obtain the sum over all $c_k(t)$ from the Smoluchowski equations
\begin{subequations}
\begin{eqnarray}
\sum_{k=1}^\infty c_k(t)&=&1-\frac t2\qquad(t\leq1)
\label{eq:36a}
\\
&=&\frac1{2t}\qquad(t\geq1)
\label{eq:36b}
\end{eqnarray}
\label{eq:36}
\end{subequations}
The limitation on the value of $t$ arises, of course, from the fact that the equation for $\sum_{k=1}^\infty c_k$
depends on the mass, which itself is only equal to 1 for $t\leq1$. 

Note that (\ref{eq:35}) is only true for $t\leq1$, so we only consider (\ref{eq:36a}), which leads to $\sigma=t/2$, and 
thus to the exact equation for the concentrations, compare with (\ref{eq:3.1}). Here we use the fact that, for the multiplicative kernel,
$\xi_j=j^{j-2}/j!$. 

We do not yet know the solution for $\sigma>1/2$, which corresponds to the post-gel phase. It is clear that the maximum
of the function (\ref{eq:33}) is now at the end of the convergence interval, that is, at $w=w_c=e^{-1}$ and $G=G_c=\frac12$
independently of $\sigma$. If we now substitute these into the r.h.s.~of (\ref{eq:20}) we obtain
\begin{equation}
c_j(t)=2\xi_je^{-j}(1-\sigma)=\frac{\xi_je^{-j}}{t},
\label{eq:37}
\end{equation}
where we have used the expression (\ref{eq:36b}) for the particle number. This is in fact the correct solution of (\ref{eq:2}) as it stands.

Note that this result is only valid for the values of $j$ which are of order one, not for those which are of order $N$: 
this is, of course, a relevant remark, since beyond the gel point there are such aggregates and they form a finite
part of the total mass. But since the asymptotic developments derived above fail for $j\sim N$, we cannot make any remarks
on the gel particles. 

This is a rather remarkable result. Indeed, for the post-gel multiplicative kernel, the Smoluchowski equations
can be meaningfully given two forms, which lead to different solutions:
\begin{subequations}
\begin{eqnarray}
\dot{c}_j(t)&=&\frac12\sum_{k=1}^{j-1}k(j-k)c_k(t)c_{j-k}(t)-jc_j(t)\sum_{k=1}^\infty kc_k(t)
\label{eq:38a}
\\
\dot{c}_j(t)&=&\frac12\sum_{k=1}^{j-1}k(j-k)c_k(t)c_{j-k}(t)-jc_j(t)
\label{eq:38b}
\end{eqnarray}
\label{eq:38}
\end{subequations}
Here (\ref{eq:38a}) is simply  (\ref{eq:2}) for $K(k,l)=kl$ and (\ref{eq:38b}) is obtained from (\ref{eq:38a})
by assuming the constancy of the mass, even though the equation's solution eventually violates it. 
The solution of (\ref{eq:38a}) is (\ref{eq:3.1}), whereas the solution of (\ref{eq:38b}) is simply (\ref{eq:3.1a})
for all times, that is, the solution is analytic. After $t=1$, of course,
the solution of (\ref{eq:38b}) no more satisfies the assumption of conserved mass,
which was used for its derivation. 	

Both these formulations are expected to be valid, each in an appropriate context. The former given by (\ref{eq:38a}) 
is known as the Stockmayer solution and is expected to hold when no interaction between the finite (sol) particles
and the gel is possible. In the opposite case, the solution described by (\ref{eq:38b}), known as the Flory solution,
is expected to hold. For a detailed discussion of this issue see \cite{ziff1980}. 

The BPA thus leads to the exact Stockmayer solution. In the case of the Marcus--Lushnikov model, 
the correct result depends on details of the model. Thus, in a related model, that of a random graph, we have initially $N$ sites and link
at each time step two randomly chosen sites. Before gelation, this is equivalent to the discrete Marcus--Lushnikov model
as $N\to\infty$. After gelation, the random graph model is known to follow the Flory solution \cite{ziff1980}. However, the random graph model's 
equivalence to the discrete  Marcus--Lushnikov model does not hold after the gel time: indeed, in 
the random graph model, it occurs with finite probability that two sites both belonging to the infinite cluster are chosen and joined, thereby 
leaving the cluster size distribution unmodified. Such a process has no equivalent in the discrete Marcus--Lushnikov model. A 
variant of the random graph model in which cycles are forbidden was analysed in \cite{ziff1980}. There it was shown that 
the decay of $c_j(t)$ as $t\to\infty$ was exponential, thereby contradicting the Stockmayer solution. 

The BPA thus rather remarkably yields exactly a well-known gelling solution. With respect to the 
discrete Marcus--Lushnikov model, it appears most likely that the solution obtained via the BPA
does not describe the model correctly, but the coincidence with a correct solution of a version of the Smoluchowski's
equations remains striking. 

For the other two kernels, we have {\color{black} the well-known relations}
\begin{equation}
Ge^{-G}=w
\label{eq:39}
\end{equation}
for the additive kernel, and
\begin{equation}
G=\frac{w}{2-w}
\label{eq:40}
\end{equation}
for the constant kernel, from which the exact solutions can be derived in an elementary manner. The fact that these solutions 
correspond to the exact solutions of (\ref{eq:2}) is readily confirmed. 

\section{Conclusions}
\label{sec:conclusions}
Summarising, we have derived analytically the solution to the Smoluchowski equations that arises from the solution of the 
discrete Marcus--Lushnikov model proposed in \cite{BPC1,BPC2,BPC3,BPC4,BPC5}. As is readily verified, in the general case
these expressions do not satisfy the properties known from the general scaling theory of the Smoluchowski equations as described in 
\cite{ley03}. Specifically, the concentration profile thus obtained agrees qualitatively well in the limit of cluster sizes large with 
respect to the typical size, but deviates strongly from the known behaviour in the opposite limit.

As is well-known, however, the three exactly solvable cases, the constant, additive, and multiplicative kernels, have the remarkable
property that
their small-size and large-size behaviours coincide. In that case, it is verified that the analytic solution derived in this paper from
the BPA does indeed coincide exactly with the corresponding solution of the Smoluchowski equation, with the exception of the
post-gel multiplicative kernel, for which the solution obtained via the BPA is the Stockmayer result, whereas the exact result
is presumably the Flory solution. 

To complicate matters further, it turns out that the BPA for the full probability distribution is almost certainly exact for all $N$
in the case of the additive and constant kernels, but that such is not the case for the multiplicative kernel. Conjecturally, one might assume
that, in the large $N$ limit, the BPA converges to the exact pre-gel solution for the full probability distribution, and similarly, that it converges to 
a probability distribution different of that of the Marcus--Lushnikov model in the post-gel case, namely one that yields the 
Stockmayer solution. Since, for small $N$, there is no sharp difference between pre-gel and post-gel stages, the existence of a 
discrepancy at small $N$ and small times is perhaps understandable. 

Considering the remarkable successes of the Bell Polynomial Ansatz, 
it would clearly be extremely desirable to obtain a better understanding of the mechanism underlying of its failure in the general case. 

\section*{acknowledgements}
Support from the grant UNAM--PAPIIT--DGAPA IN113620 and CONACyT 254515 is gratefully acknowledged.

\appendix

\section{The scaling function from (\ref{eq:20})}
\label{app:a}
In all cases except when $\nu=1$, the $\xi_j$ behave as $j^{-\lambda}$ \cite{vDongenLarge}. This implies that $G(w)$ goes as $(w_c-w)^{\lambda-1}$,
where this should be understood as meaning the singular part only. Thus, constant and linear terms dominating this behaviour 
may exist, depending on the value of $\lambda$. 

Consider first $\lambda<1$, that is, the non-gelling case. We then have $G(w)\to\infty$ as $w\to w_c$. The equation 
determining $w(\sigma)$ reads
\begin{equation}
(1-\sigma)w(\sigma)\frac{G^\prime[w(\sigma)]}{G[w(\sigma)]}=1.
\label{eq:a1}
\end{equation}
Since $\lambda<1$, both $G(w)$ and $G^\prime(w)$ diverge, and their ratio goes as $(w_c-w)^{-1}$. It follows that $(w(\sigma)-w_c)$
is proportional to $1-\sigma$ as $\sigma\to1$. $G[w(\sigma)]$ then diverges as $(1-\sigma)^{\lambda-1}$. Substituting into
(\ref{eq:20}) then yields
\begin{equation}
\frac{n_j(\sigma)}{N}=const.\cdot j^{-\lambda}(1-\sigma)^{2-\lambda}\exp\left(
-const.\cdot\frac{j}{1-\sigma}
\right)
\label{eq:a2}
\end{equation}
which has the scaling form indicated in (\ref{eq:21.03}).

If $1<\lambda<2$, then $G(w_c)$ remains finite, but $G^\prime(w)$ diverges as $(1-\sigma)^{\lambda-2}$ as $\sigma\to1$. 
A similar computation then yields
\begin{equation}
\frac{n_j(\sigma)}{N}=const.\cdot j^{-\lambda}(1-\sigma)\exp\left(
-const.\cdot\frac{j}{(1-\sigma)^{1/(2-\lambda)}}
\right)
\label{eq:a3}
\end{equation}
which again confirms the scaling form (\ref{eq:21.03}). This is particularly problematic, since it implies that no gelation arises 
for $1<\lambda<2$, in contradiction to well-known exact results \cite{jeon}.

{\color{black}
\section{Derivation of the saddle-point approximation (\ref{eq:26})}
\label{app:b}
As a first step, let us show the following elementary result: let $f(x)$ be a real valued function on the real interval
$[a,b]$, which has a unique maximum at $x^*$ which is strictly inside $[a,b]$. Then
\begin{equation}
I\lim_{N\to\infty}\frac1N\ln \int_a^b\exp\left[
Nf(x)
\right]dx=f(x^*)
\label{eq:b1}
\end{equation}
In the following we denote the integral above by $I_N$. 
Without loss of generality we assume $x^*$ to be zero. 

We now express $f(x)$ as
\begin{equation}
f(x)=f(0)+f^{\prime\prime}(0)\frac{x^2}{2}+g(x)
\label{eq:b2}
\end{equation}
where $g(x)=O(x^3)$ as $x\to0$. We now separate the integral $I_N$ in two parts, $I_N=J_N+K_N$, with
\begin{subequations}
\begin{eqnarray}
J_N&=&\int_{-N^{-\alpha}}^{N^{-\alpha}}\exp\left[
Nf(x)
\right]dx,
\label{eq:b3a}\\
K_N&=&\left(
\int_a^{-N^{-\alpha}}+\int_{N^{-\alpha}}^b
\right)
\exp\left[
Nf(x)
\right]dx,
\label{eq:b3b}
\end{eqnarray}
\label{eq:b3}
\end{subequations}
where $1/3<\alpha<1/2$. 

We shall show that $J_N$ satisfies (\ref{eq:b1}) and $K_N/J_N\to0$ as $N\to\infty$, thereby proving
the result. Since $\alpha>1/3$, we find that $Ng(x)\to0$ on the whole range of integration of $J_N$,
so that
\begin{equation}
J_N=e^{Nf(0)}\int_{-N^{1/2-\alpha}}^{N^{1/2-\alpha}}\exp\left[
-N|f^{\prime\prime}(0)|x^2/2
\right]dx\simeq e^{Nf(0)}\sqrt{\frac{2\pi}{|f^{\prime\prime}(0)|N}},
\label{eq:b4}
\end{equation}
where the final approximate equality follows from $\alpha<1/2$ and $N\gg1$. Note that, since $f(0)$ is a maximum, $f^{\prime\prime}(0)<0$,
explaining the notation using absolute values. Of course, the multiplicative term of order $N^{-1/2}$ disappears upon evaluating the 
left-hand side of (\ref{eq:b1})

On the other hand, at the border between the integration domains of $J_N$ and $K_N$, the integrand can still be evaluated using
(\ref{eq:b2}) neglecting $g(x)$. $K_N/J_N$ is therefore negligible, since it is of the order $\exp(-N^{1-2\alpha}|f^{\prime\prime}(0)|)$. Since the 
maximum is unique, the part of the integral $K_N$ for which the representation (\ref{eq:b2}) cannot be used is also exponentially small 
against $J_N$ and also negligible. 

We now need to cast the integral
\begin{equation}
I(N,S)=\frac{1}{2\pi i}\oint_C \frac{G(w|\underline{z})^{N-S}}{w^{N+1}}dw
\label{eq:b5}
\end{equation}
in the form (\ref{eq:b1}). In the following, we shall use mainly the fact that the power series of $G(w|\underline{z})$
in terms of $w$ has only positive coefficients, and that these asymptotically behave as a power-law $z_k\sim k^{-\lambda}$.

Defining $\sigma=S/N$, we have
\begin{equation}
I(N,N\sigma)=\frac{1}{2\pi i}\oint_C \exp\left[N\big((1-\sigma)\ln G(w|\underline{z})-\ln w\big)\right]\frac{dw}{w}.
\label{eq:b6}
\end{equation}
We define the abbreviation:
\begin{equation}
H(w|\underline{z})=(1-\sigma)\ln G(w|\underline{z})-\ln w
\label{eq:b7}
\end{equation}
The quantity $w^*(\sigma|\underline{z})$ corresponds to the minimum of the function $H(w|\underline{z})$ on the real interval
$[0,w_c(\underline{z})]$. In the following, we shall use as a contour $C$ the circle of radius $w^*(\sigma|\underline{z})$.

Now, since $G(w|\underline{z})$ has only positive coefficients, by a standard theorem of complex analysis \cite{ahl}, its maximum
modulus on $C$ lies on the positive real axis, and the same holds for $G(w|\underline{z})/w^{1/(1-\sigma)}$, since $|w|$ is constant on $C$.
The same thus holds as well for the logarithm and hence for the real part of $H(w|\underline{z})$. Therefore the only maximum on $C$
of $\operatorname{Re} H(w|\underline{z})|$ is at $w^*(\sigma|\underline{z})$. 

Note however that, in order additionally to ensure that the maximum modulus arising on the real axis is the only one on $C$, we must 
exclude such irregular growth as might occur, for instance, if the coefficients of the power series of $G$ are only non-zero for even values of
$k$. We therefore implicitly assume that such behaviour does not arise. 

We therefore understand the behavior of the modulus of $H$. Let us now look at its real and imaginary parts. The real part coincides withe the
function itself on the real axis and is minimum at  $w^*(\sigma|\underline{z})$ with respect to purely real variations. 
From the Cauchy--Riemann differential equations \cite{ahl} one finds that the imaginary part also has vanishing derivative with respect to
imaginary variations. Thus, since the contour $C$ is vertical at $w^*(\sigma|\underline{z})$, the imaginary part has zero derivative along $C$.
But since $H$ is real on the real axis, by the Schwarz reflection principle \cite{ahl}, the imaginary part of $H$ is odd, so that it grows
cubically (or possibly faster) as the distance to the real axis, and can thus be neglected in the immediate vicinity of the point  $w^*(\sigma|\underline{z})$.
We may now apply (\ref{eq:b1}) without problems and obtain the desired result (\ref{eq:26}).

Finally, it should be pointed out that the above approximation describes the $z_s$-dependence of the integral (\ref{eq:b6}) only for $s\ll N$.
Indeed, $G(w|\underline{z})$ in can be replaced without changing the results by $G_N(w|\underline{z})$ given by
\begin{equation}
G_N(w|\underline{z})=\sum_{j=1}^N\xi_jz_jw^j.
\end{equation}
In particular, it follows that such integrals as (\ref{eq:b6}) are {\em independent\/} of $z_s$ for $s>N$ and the above approximate
approach does not apply for $s$ of order $N$.
}

\end{document}